\documentclass[aps,pre,twocolumn,superscriptaddress,showpacs]{revtex4}
\usepackage{amssymb}
\usepackage{epsfig}
\usepackage{amsmath}
\usepackage{times}
\usepackage[T1]{fontenc}
\usepackage{amsmath}

\begin{document}

\title{Geometric properties of the additional third-order transitions in the two-dimensional Potts model}
\author{Wei Liu}
\email{weiliu@xust.edu.cn}
\affiliation{College of Sciences, Xi'an University of Science and Technology, Xi'an, 710054, P. R. China}
\author{Xin Zhang}
\affiliation{College of Sciences, Xi'an University of Science and Technology, Xi'an, 710054, P. R. China}
\author{Lei Shi}
\affiliation{College of Sciences, Xi'an University of Science and Technology, Xi'an, 710054, P. R. China}
\author{Kai Qi}
\email{kqi@mail.sim.ac.cn}
\affiliation{2020 X-Lab, Shanghai Institute of Microsystem and Information Technology, Chinese Academy of Sciences, ShangHai, 519085, P. R. China}
\author{Xiang Li}
\email{lixiang11@alumni.nudt.edu.cn}
\affiliation{National Innovation Institute of Defense Technology, PLA Academy of Military Science, Beijing, 100071, P. R. China}
\author{Fangfang Wang}
\affiliation{School of Systems Science, Beijing Normal University, Beijing, 100875, P. R. China}
\affiliation{International Academic Center of Complex Systems, Beijing Normal University, ZhuHai, 519087, P. R. China}
\affiliation{Department of Systems Science, Faculty of Arts and Sciences, Beijing Normal University, ZhuHai, 519087, P. R. China}
\author{Zengru Di}
\affiliation{School of Systems Science, Beijing Normal University, Beijing, 100875, P. R. China}
\affiliation{International Academic Center of Complex Systems, Beijing Normal University, ZhuHai, 519087, P. R. China}
\affiliation{Department of Systems Science, Faculty of Arts and Sciences, Beijing Normal University, ZhuHai, 519087, P. R. China}

\date{\today}

\begin{abstract}
Within the canonical ensemble framework, this paper investigates the presence of higher-order transition signals in the $q$-state Potts model (for $q \geq 3$), using two geometric order parameters: isolated spins number and the average perimeter of clusters. Our results confirm that higher-order transitions exist in the Potts model, where the number of isolated spins reliably indicates third-order independent transitions. This signal persists regardless of the system's phase transition order, even at higher values of $q$. In contrast, the average perimeter of clusters, used as an order parameter for detecting third-order dependent transitions, shows that for $q = 6$ and $q = 8$, the signal for third-order dependent transitions disappears, indicating its absence in systems undergoing first-order transitions. These findings are consistent with results from microcanonical inflection-point analysis, further validating the robustness of this approach.

\end{abstract}


\maketitle
\section{Introduction}

The Potts model \cite{potts1952some,wu1984potts} generalizes the Ising model to multiple states \cite{onsager1944crystal,brush1967history}, with different spin states represented by integer values of $q$. It has become an important framework in statistical mechanics, contributing significantly to the understanding of phase transitions and critical phenomena in complex systems. Originally developed to explore magnetic properties \cite{wu1984potts}, the Potts model has found applications across diverse fields, including optimizing communication networks like Wireless Body Area Networks and MESH networks \cite{paszkiewicz2020responsiveness}. Recent studies have expanded its scope to disciplines such as biophysics and statistical physics. For instance, Bae and Tai \cite{tai2023potts} introduced a four-phase Potts model for image segmentation, while Morcos et al. \cite{morcos2011direct} applied it to protein structure prediction using direct coupling analysis. The model's relevance has recently been extended to machine learning, where Rende and colleagues \cite{rende2024mapping} mapped the self-attention mechanism onto a generalized Potts model, demonstrating its effectiveness in solving the inverse Potts problem. These developments highlight the model's expanding interdisciplinary impact.

Traditional statistical physics and thermodynamics have been highly successful in explaining the behavior of phase transitions in systems. A phase transition is a cooperative phenomenon involving multiple factors such as temperature and pressure. The structure and properties of a system evolve in response to variations in certain order parameters, primarily temperature \cite{phaseconcept}. In the $q$-state Potts model, for \( 1 \leq q \leq 4 \), the phase transition of the Potts model is continuous \cite{duminil2017continuity,baxter1973potts}, while for \( q > 4 \), the phase transition becomes discontinuous \cite{baxter1973potts,duminil2021discontinuity,igloi1999boundary,hamer1981q,igloi1983first,loureiro2010curvature,loureiro2012geometrical}. By incorporating bond percolation theory \cite{fortuin1972random,stephen1976percolation,wu1978percolation,lee2005scale}, we can easily determine whether a phase transition is continuous or not. Percolation theory \cite{stauffer2018introduction,grimmett1999inequalities,essam1980percolation,saberi2015recent} is the natural framework to study the properties of cluster-like structures of a system. Extensive research on percolation, especially focusing on both bond and site percolation in two and three dimensions, has been conducted, with significant contributions from Youjin Deng and his collaborators, uncovering phase transitions in percolation models \cite{wang2013bond,feng2008percolation,deng2003simultaneous,deng2005monte,xu2014simultaneous}.
As we know, the early warning of phase transitions hold theoretical significance and practical value \cite{scheffer2009early,freeman2012helf}. It raises the question of whether there exist typical and universal behaviors that can act as precursors to main phase transitions, and whether these have the potential to be applied beyond classical magnetic systems, such as artificial swarms \cite{bechinger2016active} or active bacterial colonies \cite{copeland2009bacterial, qi2022emergence}.

Over the past few decades, microcanonical analysis has emerged as a valuable approach for identifying phase transitions in physical systems \cite{gross2001microcanonical,junghans2006microcanonical}. Microcanonical inflection-point analysis (MIPA) has been recognized as an effective method for studying phase transitions in finite-size systems. Qi and Bachmann \cite{qi2018classification} expanded this approach to effectively identify higher-order transitions, distinguishing between independent and dependent transitions. The inflection point in a higher-order derivative of the microcanonical entropy specifically marks the point at which the monotonicity of the function changes, signaling the occurrence of a higher-order transition. Independent transitions, akin to conventional phase transitions, occur independently of other cooperative processes within the system. Dependent transitions, in contrast, are contingent upon the occurrence of lower-order transitions and take place at higher energy levels, representing higher-order phenomena.

Utilizing the exact density of states (DOS) of the Ising model \cite{beale1996exact}, Sitarachu performed a comprehensive analysis of 1D and 2D finite-size Ising models, identifying higher-order transitions in the 2D system \cite{sitarachu2020exact}. By using geometric order parameters in the canonical ensemble, Sitarachu et al. effectively demonstrated the correspondence between canonical and microcanonical ensemble methods for identifying third-order transitions \cite{sitarachu2022evidence}. Their analysis identified two forms of third-order transitions: independent and dependent transitions, and determined how to use geometric order parameters to locate the positions of these transitions. The third-order independent transition occurs before the phase transition and is geometrically associated with the peak in the number of isolated spins, which disrupts the order of the system. This maximum is observed below the transition temperature. In contrast, the third-order dependent transition emerges beyond the transition point and is associated with a local minimum in the first derivative of the average cluster size with respect to temperature, indicating a shift in the decay rate of clusters in the disordered phase. This dependent transition is marked by a minimum above the transition temperature.

An important question arises: are these geometric signatures of third-order transitions universally applicable across different models? To address this, it becomes essential to test these methods on another model. This paper aims to explore the presence of third-order transition signals in the $q$-state Potts model (for $q \geq 3$). Through simulations in the canonical ensemble, we observed that applying the same geometric order parameters used in the Ising model to identify third-order transitions in the Potts model is not straightforward. In this work, we have redefined isolated spins and substituted the average cluster size with the average cluster perimeter. The rationale behind these modifications will be elaborated upon in the subsequent sections.

This paper is organized as follows: Section II provides an overview of the Potts model and the Swendsen-Wang algorithm, along with an analysis of the third-order transitions identified using geometric order parameters. Section III examines the main phase transition order for different values of $q$ and investigates the presence of higher-order transitions, emphasizing their dependence on the phase transition type. Finally, Section IV summarizes the article.

\section{The model and method}

\subsection{Potts Model}
The $q$-state Potts model \cite{wu1982potts}, a generalization of the Ising model, is extensively utilized to study phase transitions in statistical mechanics. In this model, spins on a lattice may assume one of $q$ discrete states, as opposed to the binary states in the Ising model. There is no external magnetic field in this model. The energy for a specific configuration of spins $X=(s_1, s_2, \dots, s_N)$, on a lattice with $N=L \times L$ sites, is expressed as:
\begin{equation}
E(X) = -J \sum_{\langle i,j \rangle} \delta(s_i, s_j)
\end{equation}
where $s_i$ represents the spin at site $i$, which may take integer values from $0$ to $q-1$, and $\delta(s_i, s_j)$ is the Kronecker delta function, which equals $1$ when $s_i = s_j$ and $0$ otherwise. The summation is performed over all nearest-neighbor pairs $\langle i,j \rangle$, and $J > 0$ is the coupling constant that favors alignment of neighboring spins, reflecting ferromagnetic interactions. When $q=2$, the Potts model reduces to the Ising model, while larger values of $q$ lead to increasingly complex behavior and phase transition dynamics. The model has been thoroughly investigated for its applications in understanding critical phenomena and phase transitions \cite{baxter1973potts}.

\subsection{Swendsen-Wang Algorithm}
It is well established \cite{fortuin1972random} that the Potts model is intricately connected to problems of connectivity and percolation in graph theory. The Fortuin-Kasteleyn transformation \cite{kasteleyn1969phase} enables a mapping of the original model, which suffers from critical slowdown, into one where such slowdown effects are significantly mitigated. The transformation places a bond between each pair of interacting Potts spins on the lattice with the probability:
\begin{equation}
p = 1 - e^{-K \delta_{\sigma_i, \sigma_j}},
\label{equationpro}
\end{equation}
where \( K = {J}/{k_B T} \), and \( J = 1 \). A bond is placed with probability \( p \) only when \( \sigma_i = \sigma_j \), indicating that the spins of the nearest neighbors are in the same state. This implies that bonds are only formed with a non-zero only when the probability when the corresponding pair of spins on the original lattice are in the same state. The process must be repeated for all pairs of spins, resulting in a lattice where bonds connect certain sites, forming clusters of varying sizes and shapes.

Swendsen and Wang \cite{swendsen1987nonuniversal} successfully applied the Fortuin–Kasteleyn transformation in Monte Carlo simulations. The process involves traversing the lattice and placing bonds between each pair of spins with the probability given by Equation (\ref{equationpro}). The Hoshen–Kopelman \cite{hoshen1976percolation} method is then used to identify all clusters of sites connected by the bond network. Subsequently, each cluster is assigned a new spin value from the set \( q = \{0,1,2,\dots,q-1\} \), ensuring that all sites within a given cluster share the same randomly chosen spin value. This constitutes one Monte Carlo step in our algorithm. In the next Monte Carlo step, based on the spin configuration from the previous step, new bonds are re-established between neighboring spins according to the predefined probability. The Hoshen-Kopelman algorithm \cite{hoshen1976percolation} is once again used to identify clusters, after each cluster is reassigned a new spin value uniformly from the set \( q = \{0,1,2,\dots,q-1\} \). This process repeats iteratively until the desired number of Monte Carlo steps is reached. It is important to note that in our algorithm, a dedicated array is used to store the bond states between spins. This array is reset at the start of each Monte Carlo step, ensuring that the bonding configuration is entirely refreshed, with no correlation to the bonds in the previous step.

In our algorithm, we emphasize that during each Monte Carlo (MC) step, the Hoshen-Kopelman (H-K) algorithm is applied twice for cluster statistics: first to identify clusters within the system, followed by a spin-flipping process in which F-K clusters are defined by introducing a bond between two nearest-neighbor spins with identical orientations, based on the probability outlined in equation (\ref{equationpro}). However, once the F-K clusters are established, it is essential to ascertain the true physical state of the system to compute the order parameters accurately. This requires the formation of clusters solely from nearest-neighbor spins that share the same orientation after the spin-flipping operations, ensuring that the final cluster information aligns with the actual physical system. Thus, while the construction of F-K clusters addresses the issue of critical slowdown during spin flips, the accurate computation of physical quantities in the system necessitates considering the actual physical clusters present.

Since the probability of placing a bond between pairs of sites depends on temperature, it is clear that the resultant cluster distributions will vary dramatically with temperature. At very high temperatures, the clusters tend to be quite small. At very low temperatures, virtually all sites with closest neighbors in the same state will end up in the same cluster, leading to oscillations between similar structures. Near the transition point, a diverse array of clusters emerges, resulting in each configuration differing significantly from its predecessor. This effectively reduces the phenomenon of critical slowdown. Using the Swendsen-Wang algorithm (S-W), critical slowdown is significantly mitigated, facilitating faster convergence and more efficient sampling in Monte Carlo simulations \cite{newman1999monte}.

If we aim to study the behavior of phase transitions, the F-K transformation is an effective tool for this purpose. In the F-K clusters, the probability of a randomly selected spin belonging to the largest cluster, denoted as \( \langle P_{\infty} \rangle = \langle n_{\infty} \rangle / L^2 \), where \( \langle n_{\infty} \rangle \) represents the size of the largest cluster, exhibits behavior consistent with the magnetization \( \langle M \rangle \) calculated from real clusters for temperatures \( T \leq T_{\text{c}} \). However, for \( T > T_{\text{c}} \), the value of \( \langle P_{\infty} \rangle \) is slightly smaller than that of \( \langle M \rangle  \) \cite{de1990monte}. In the paramagnetic phase (\( T > T_{\text{c}} \)), discrepancies arise between these two quantities, as many clusters contribute to the magnetization \( \langle M \rangle \), rather than solely the largest cluster, which is represented by \( \langle P_{\infty} \rangle \). Fluctuation quantities, such as specific heat and susceptibility, exhibit related differences that arise from distinct contributions of clusters. In particular, contributions must be separated from clusters smaller than the largest and those from the size of the largest cluster itself \cite{landau2021guide}. To identify additional signals of higher-order phase transitions, especially in the paramagnetic phase, it is crucial to seek an alternative and appropriate order parameter that can more effectively capture subtle changes associated with these transitions \cite{sitarachu2022evidence}.

\subsection{Third-order transitions}

\begin{figure}[htbp]
  \centering
  \includegraphics[width=0.8\columnwidth]{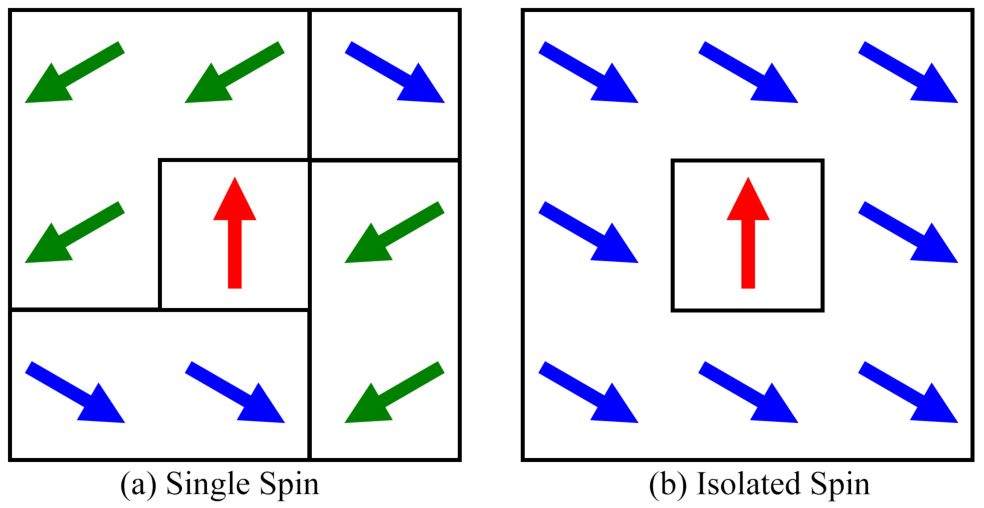}
  \caption{\label{fig:sinVSiso} Comparison of Single Spins and Isolated Spins in the $q = 3$ state Potts model. Red represents $q = 0$, blue represents $q = 1$, and green represents $q = 2$. The figure illustrates the distinction between single spins, which differ from their nearest neighbors regardless of cluster membership, and isolated spins, which differ in orientation from their four nearest neighbors, all of which belong to the same cluster.}
\end{figure}

Microcanonical inflection-point analysis demonstrates the occurrence of a third-order independent transition prior to the phase transition temperature $T_{\text{c}}$. Geometrically, this third-order independent transition corresponds to a peak in the count of isolated spins, observed at a temperature \( T_{\text{ind}} \) lower than the phase transition temperature \( T_{\text{c}} \). In the Ising model, the third-order transition is observed at $T_{\text{ind}} \approx 2.229$ \cite{sitarachu2022evidence}. However, when the same definition of isolated spins from the Ising model is applied to the Potts model, the number of isolated spins increases monotonically with temperature, without showing evidence of a third-order independent transition. We speculate that this may be because, in the Potts model in the $q$ state (for $q \geq 3$), the spins can adopt more than two states. If isolated spins are simply defined as those with a spin direction different from that of the four nearest neighbors, then for the \( q \)-state Potts model with \( q \geq 3 \), the system will contain two types of single-spin clusters: one as shown in Fig.~\ref{fig:sinVSiso} (a) and the other as shown in Fig.~\ref{fig:sinVSiso} (b). For the type of single-spin shown in Fig.~\ref{fig:sinVSiso} (a), the ordered state of the region it resides in has already been broken. Therefore, it is meaningless to include it in the statistics. We only need to consider the isolated spins as shown in Fig.~\ref{fig:sinVSiso} (b).

As a result, in the $q$-state Potts model (for $q \geq 3$), it is necessary to redefine the concept of isolated spins. In the Ising model, it is observed that, due to the two-state nature of the Ising model, an isolated spin is surrounded by four nearest neighbors that have the same orientation, which is opposite to that of the spin itself. This leads to the hypothesis that the role of an isolated spin is to serve as a disrupter of order within the system. Extending this logic to the $q$-state Potts model, isolated spins are similarly defined. In an ordered cluster, an isolated spin serves as a disrupter, with a different orientation from its four nearest neighbors, as shown in Fig.~\ref{fig:sinVSiso} (b), all of which belong to the same cluster. The isolated spin thus behaves like a "nail" driven into an ordered cluster, demonstrating its function as a source of disorder within an otherwise ordered system.

\begin{figure}[ht]
    \centering             
    \includegraphics[width=0.3\textwidth]{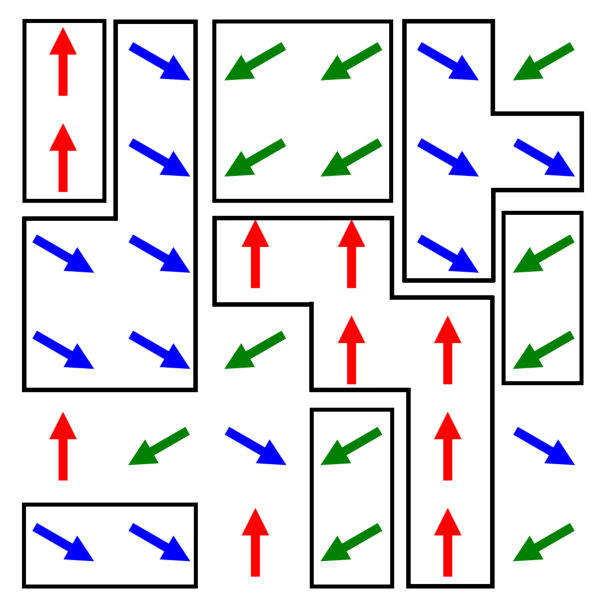}
    \caption{Cluster perimeter illustration for the $q=3$ Potts model. This figure demonstrates the relationship between cluster size and perimeter in the $q=3$ Potts model. While the cluster size only indicates the number of spins within a cluster, the perimeter more accurately captures the boundary characteristics of the cluster. Clusters with the same number of spins can exhibit different configurations, resulting in distinct perimeters. This distinction makes the perimeter a more precise and effective metric for analyzing cluster configurations and identifying signals of higher-order transitions.
}
    \label{fig:spin_lattice_no_bonds}
\end{figure}

Microcanonical inflection-point analysis demonstrates that an additional third-order dependent transition occurs within the paramagnetic phase, at a temperature \( T_{\text{de}} \) distinct from the phase transition temperature \( T_{\text{c}} \), where \( T_{\text{de}} \) is larger than \( T_{\text{c}} \). In the canonical ensemble, this corresponds to a slight shift in the rate at which the average cluster size decreases after the critical phase transition in the Ising model, specifically when $T > T_{\text{c}}$. Mathematically, this is identified by a local minimum in the first derivative of the average cluster size with respect to temperature. The temperature at which this minimum occurs indicates the occurrence of the third-order dependent transition, which, in the Ising model, is found to be $T_{\text{de}} \approx 2.567$ \cite{sitarachu2022evidence}. Following this logic, this approach is extended to the \( q \)-state Potts model (for \( q \geq 3 \)) to search for the signal of a third-order dependent transition, but the attempt did not yield any successful results.

Subsequently, several order parameters were investigated in the $q$-state Potts model (for $q \geq 3$), and ultimately, the average perimeter of clusters was identified as a indicator for detecting the third-order dependent transition in the model. The process of identifying the average perimeter as an order parameter was somewhat unexpected. After realizing that the average cluster size could not pinpoint the third-order dependent transition in the $q$-state Potts model, an analysis of this parameter was conducted in detail. The average cluster size represents the average area of clusters, indicating the number of spins within each cluster. Since area failed to capture the signal, the focus was shifted to examining the average perimeter, which characterizes the number of spins on the boundary of each cluster,as shown in Fig.~\ref{fig:spin_lattice_no_bonds}. Interestingly, during the analysis of the temperature dependence of the average perimeter, a similar trend was observed to that of the average cluster size in the 2D Ising model.

After identifying the average perimeter $G$ as the order parameter for detecting the third-order dependent transition, the precise definition of the average perimeter, $\langle G \rangle$, is provided. $G$ can be defined as the average perimeter of clusters that contain more than one spin in a given spin configuration $X$:
\begin{equation}
G = \frac{1}{n} \sum_{l} P_l
\end{equation}
where $l$ labels the clusters with more than one spin, $P_l$ is the perimeter of cluster $l$, and $n$ is the total number of clusters with more than one spin in $X$. The statistical average is then obtained as:
\begin{equation}
\langle G \rangle = \frac{1}{Z} \sum_{X} G(X) e^{-E(X)/k_B T},
\end{equation}
where $T$ is the canonical temperature and $Z$ is the partition function defined as
\begin{equation}
Z = \sum_{X} \exp\left(-\frac{E(X)}{k_B T}\right).
\end{equation}

\section{Results and discussion}

The following section discusses the results obtained from Swendsen-Wang spin cluster simulations and cluster analysis to elucidate the system behavior associated with the additional transitions in the Potts model, as identified by microcanonical inflection-point analysis.

\subsection{Phase transition}

\begin{figure}
	\epsfig{figure=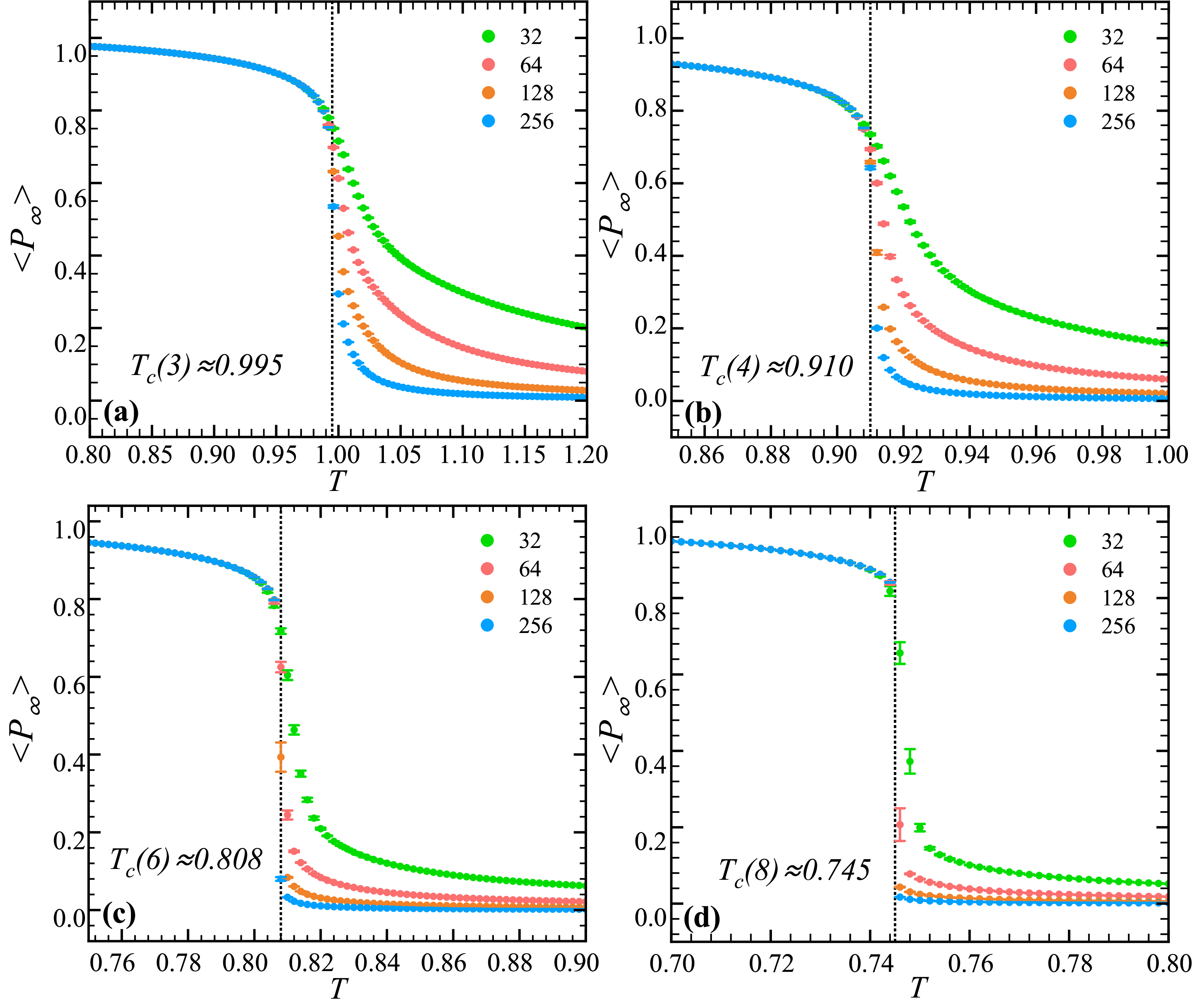,width=1.\linewidth} \caption{
		The variation of $\langle P_{\infty} \rangle$ with temperature is shown. The black dashed line marks the precise solution of the system's phase transition temperature \(T_{\text{c}}\).As shown in panel (a) and (b), it is observed that at $T_{\text{c}}$, the $q=3, 4$ Potts models exhibit characteristics of a continuous phase transition, indicating that the system undergoes a second-order phase transition for $q=3, 4$. In contrast, in panel (c) and (d), for $q=6, 8$, $\langle P_{\infty} \rangle$ shows a discontinuity at $T_{\text{c}}$, signifying that the system undergoes a first-order phase transition for $q=6, 8$.
	}
	\label{fig:LCS}
\end{figure}

The primary goal of this paper is to determine the locations of both the third-order independent and dependent transitions in the $q$-state Potts model (for $q \geq 3$) for finite system sizes. The system sizes studied in this work are based on square lattices with edge lengths $L = 32, 64, 128, 256$. As derived by Wu \cite{wu1982potts}, the exact solution for the phase transition temperature $T_{\text{c}}$ of the $q$-state Potts model (for $q \geq 3$) is given by:
\begin{equation}
T_{\text{c}} = \frac{1}{\ln(1 + \sqrt{q})}
\end{equation}

Therefore, for $q = 3, 4, 6, 8$, the corresponding phase transition temperatures approximate $T_{\text{c}}(3) \approx 0.995$, $T_{\text{c}}(4) \approx 0.910$, $T_{\text{c}}(6) \approx 0.808$, and $T_{\text{c}}(8) \approx 0.745$, respectively.

After identifying all the clusters in the system, various types of clusters emerge, and it is inevitable that the system contains a largest spin cluster. The phase behavior of $\langle P_{\infty}\rangle$ closely mirrors that of $\langle M \rangle$ for $T \leq T_{\text{c}}$, with minor deviations observed for $T > T_{\text{c}}$, while the determination of whether the phase transition is continuous remains largely consistent \cite{kasteleyn1969phase}. Consequently, analyzing the phase transition behavior of $\langle P_{\infty}\rangle$ facilitates identifying the order of the phase transition in the $q$-state Potts model (for $q \geq 3$) \cite{baxter2016exactly}. In Fig.~\ref{fig:LCS} (a) and (b), it is observed that for $q=3$ and $q=4$, $\langle P_{\infty}\rangle$ changes continuously with temperature, which aligns with the results from Wu \cite{wu1982potts} and Baxter \cite{baxter1973potts}. This confirms that the phase transitions in the $q$-state Potts model for $q=3$ and $q=4$ are critical, i.e., second-order phase transitions. For $q=6$ and $q=8$, $\langle P_{\infty}\rangle$ exhibits discontinuous behavior, consistent with the numerical simulation results reported in the existing literature \cite{baxter1973potts,igloi1999boundary,hamer1981q,igloi1983first,loureiro2010curvature,loureiro2012geometrical}, indicating a first-order phase transition when $q=6$ and $q=8$.

\begin{figure}
	\epsfig{figure=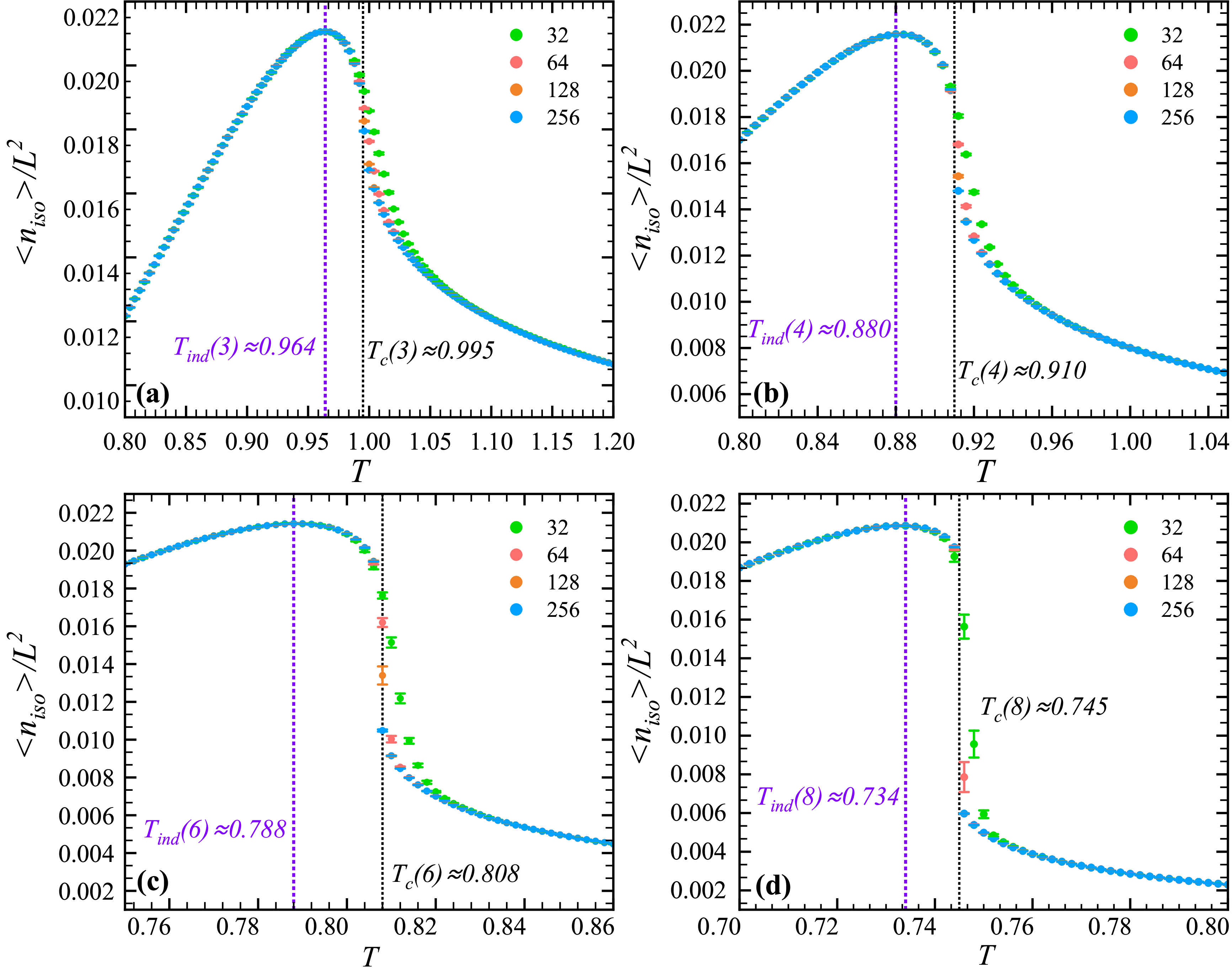,width=1.\linewidth}
    \caption{
		The variation of isolated spins (ISO) with temperature for $q=3, 4, 6, 8$ is depicted in panels (a)-(d), respectively. The purple dashed line marks the temperature corresponding to the third-order independent transition, while the black dashed line indicates the precise solution of the system's phase transition temperature \(T_{\text{c}}\). As shown in the figure, the temperatures corresponding to the third-order independent transition are identified as $T_{\text{ind}}(3) = 0.964$, $T_{\text{ind}}(4) = 0.880$, $T_{\text{ind}}(6) = 0.788$, and $T_{\text{ind}}(8) = 0.734$. 
	}
	\label{fig:ISO}
\end{figure}

\subsection{Third-order independent transition in the ferromagnetic phase}

As shown in Fig.~\ref{fig:ISO} (a)-(d), the number of isolated spins in the $q = 3,4,6,8$ Potts models, respectively, varies as a function of temperature. It is observed that the number of isolated spins initially increases with temperature, reaching a peak at $T_{\text{ind}}(3) \approx 0.964$ for the 3-state Potts model, $T_{\text{ind}}(4) \approx 0.880$ for the 4-state Potts model, $T_{\text{ind}}(6) \approx 0.788$ for the 6-state Potts model, and $T_{\text{ind}}(8) \approx 0.734$ for the 8-state Potts model. In Fig.~\ref{fig:ISO}, comparing the temperature of the third-order independent transition, $T_\text{ind}$, with the phase transition temperature, $T_{\text{c}}$, at the same $q$, the differences are approximately $\Delta T(3) \approx 0.031$, $\Delta T(4) \approx 0.030$, $\Delta T(6) \approx 0.020$, and $\Delta T(8) \approx 0.011$. It is evident that for $q=3$ and $q=4$, the third-order independent transition occurs at a notable distance from the phase transition, with $\Delta T(3) \approx \Delta T(4)$. However, for $q=6$ and $q=8$, $\Delta T$ starts to decrease, and as $q$ increases, $\Delta T$ becomes smaller. Although the temperatures of the third-order independent transition and the system's phase transition become closer, microcanonical inflection point analysis shows a re-entrant behavior of the inverse temperature as energy increases \cite{wang2024exploring}, meaning that a single temperature corresponds to two energy levels. Near $T=T_{\text{c}}$, the system exhibits a two-phase coexistence. Therefore, despite the proximity of these temperatures, there remains a distinct difference in the system's states.

\begin{figure*}
	\epsfig{figure=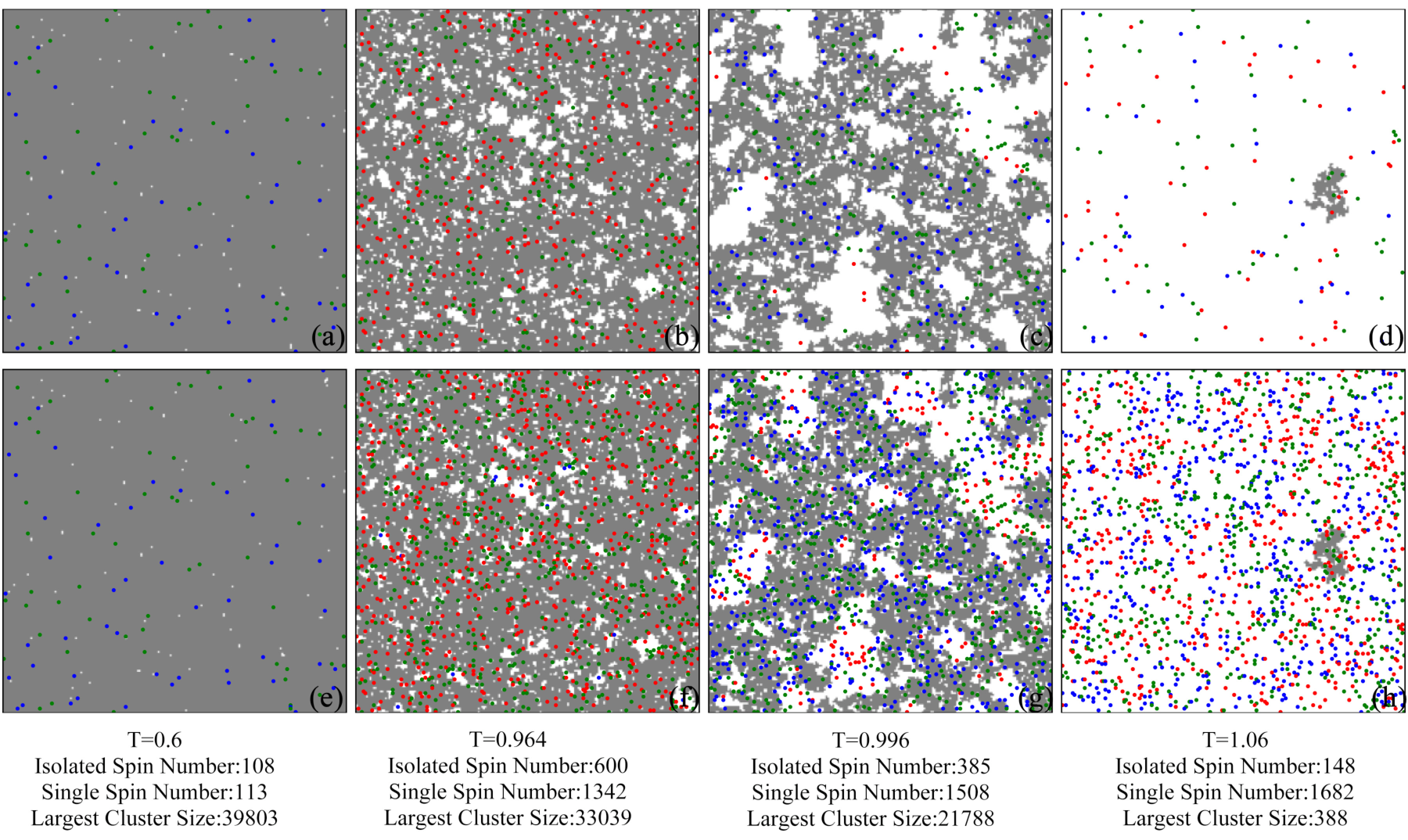,width=1.\linewidth} 
	\caption{
		A comparison of isolated spins (ISO) and single spins in the $q=3$ Potts model is provided on a lattice with $N = 200 \times 200$. In the figure, each column compares isolated spins and single spins at the same temperature and configuration. The largest cluster size (LCS) in the system is shaded in gray, while all other clusters containing more than two spins are depicted in white, regardless of their spin orientations. The red points represent spins with $s_i = 0$, the green points represent $s_i = 1$, and the blue points represent $s_i = 2$.
	}
	\label{fig:spin03_configuration}
\end{figure*}

As shown in Fig.~\ref{fig:spin03_configuration}, it becomes apparent that at low temperatures ($T < T_{\text{c}}$), the largest cluster dominates the system, nearly occupying the entire space, with little variation in its size and configuration. Prior to the phase transition, the percolation cluster continues to dominate, oscillating between similar structures. Disrupting the largest cluster at low temperatures can be achieved by isolated spins with orientations different from those within the largest cluster. During the spin-flip process, the largest cluster has a probability of being flipped, resulting in changes to the required isolated spin states,as shown in Fig.~\ref{fig:spin03_configuration} (a)-(b) and (e)-(f). As the temperature increases, the size of the largest cluster decreases. At the phase transition temperature \( T_{\text{c}} \), its size can only maintain a span across the entire lattice, but its dominance is much weaker compared to the low-temperature state. As shown in Fig.~\ref{fig:spin03_configuration} (c) and (g), with further increase in temperature, the largest cluster begins to collapse rapidly. From Fig.~\ref{fig:spin03_configuration} (d) and (h), we can observe that the size of the largest cluster becomes very small, and its size is almost indistinguishable from that of other clusters in the system. In the disordered phase, there is no percolating cluster left in the system, and at this point, studying the influence of the largest cluster size on the system in the disordered phase becomes meaningless. In the paramagnetic phase at high temperatures, the system becomes highly disordered, forming many small clusters and a large number of single spins. Panels (a)-(d) show the variation of isolated spins with temperature, while panels (e)-(f) illustrate the changes in single spin clusters with temperature. The number of isolated spins reaches a maximum at $T_{\text{ind}}$ and then decreases as temperature increases further, while the number of single spins rises continuously, indicating increasing disorder in the system.

For an arbitrary value of $q$, the behavior of $\langle n_{\text{iso}} \rangle / L^2$ as a function of temperature remains consistent across different system sizes $L$. This implies that for the same value of $q$, the proportion of isolated spins to the system size $L^2$ is invariant across different system sizes. Based on this, combined with Fig.~\ref{fig:spin03_configuration}, we observe that isolated spins are essentially a special type of single-spin cluster. A single spin appearing in the system can be considered a "mutant" spin. If this mutant spin emerges in an ordered region, it becomes an isolated spin. However, if it appears in a disordered region, it is merely a simple single-spin cluster. At low temperatures, as illustrated in Fig.~\ref{fig:spin03_configuration} (a) and (e), isolated spins and single-spin clusters are nearly indistinguishable. As the temperature increases, the number of isolated spins begins to grow, with some merging into new small clusters. In certain cases, isolated spins with different orientations may cluster together, at which point they are no longer considered isolated spins because they disrupt the surrounding region, rendering it disordered. Consequently, they are reclassified as simple single-spin clusters. When the system's temperature exceeds $T_{\text{ind}}$, the system's order begins to collapse rapidly. At this stage, ordered regions in the system are quickly disrupted by isolated spins, leading to an increase in the number of small clusters and single-spin clusters, while the number of isolated spins decreases sharply. This observation underscores the necessity of redefining isolated spins. By adopting this redefinition, the physical significance of isolated spins can be better explained, and their behavior before the system's transition point becomes more pronounced, enabling them to serve as precursors to the system's transition behavior.

The extremum of $\langle n_{\text{iso}} \rangle / L^2$ is examined for different values of $q$: $\langle n_{\text{iso}}(3) \rangle / L^2 \approx \langle n_{\text{iso}}(4) \rangle / L^2 \approx \langle n_{\text{iso}}(6) \rangle / L^2 \approx \langle n_{\text{iso}}(8) \rangle / L^2 \approx 0.0215$. This is reasonable because, at $T < T_{\text{c}}$, the percolation cluster remains dominant, and in a square lattice system, the size and configuration of the percolation cluster are similar across different values of $q$. Hence, the proportion of isolated spins required to break the percolation cluster is nearly the same. This also explains why the third-order independent transition does not depend on the order of the system's phase transition and is consistently present.

Finally, an explanation is provided for why the temperature associated with the third-order dependent transition decreases as $q$ increases. By referring to equation (\ref{equationpro}) and the SW algorithm, along with Fig.~\ref{fig:spin03_configuration}, it is noted that when a region is to generate an isolated spin, its spin orientation must differ from that of the percolation cluster. For smaller values of $q$, such as $q=3$, an isolated spin has only two alternative spin orientations, with the third option being reintegration into the percolation cluster. In this case, the probability that a spin becomes isolated is $2/3$, requiring more energy to induce a spin variation. However, for $q=8$, the situation is different. The probability that a spin becomes isolated increases to $7/8$, significantly reducing the energy required. Consequently, the temperature at which isolated spins reach their extremum decreases as $q$ increases.

\begin{figure*}
	\epsfig{figure=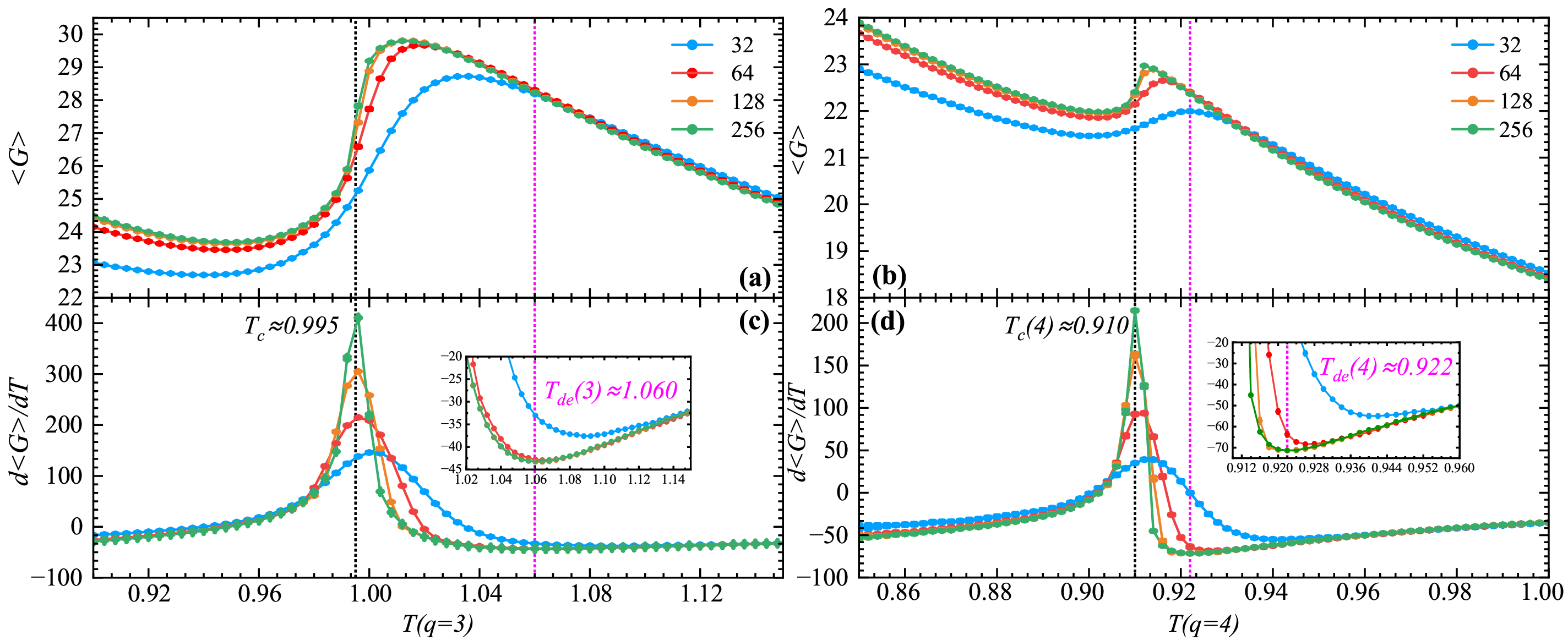,width=1.\linewidth}
    \caption{
		Panels (a) and (b) show the variation of the average perimeter $\langle G \rangle$ with temperature for $q=3, 4$, while panels (c) and (d) depict the first derivative of the average perimeter with respect to temperature, $d\langle G \rangle / dT$. The black dashed line represents the exact solution for the phase transition temperature of the corresponding $q$-state Potts model. For any value of $q$, it is observed that as $L$ increases, the location of the maximum in $d\langle G \rangle / dT$ approaches the phase transition temperature. The pink line indicates the location of the third-order dependent transition temperature, with $T_{\text{de}}(3) \approx 1.06$ and $T_{\text{de}}(4) \approx 0.922$. 
	}
	\label{fig:PEM34}
\end{figure*}

\subsection{Third-order dependent transition in the paramagnetic phase}

Fig.~\ref{fig:PEM34} illustrates the temperature dependence of the average perimeter $\langle G \rangle$ for the three-state and four-state Potts models. The first row displays the variation of the average perimeter $\langle G \rangle$ with temperature, exhibiting a backbending pattern analogous to that observed in the Ising model. In Fig.~\ref{fig:PEM34} (c) and (d), it is observed that $d\langle G \rangle / dT$ exhibits a peak, and as the system size increases, the peak of $d\langle G \rangle / dT$ approaches the system's transition temperature $T_{\text{c}}$, for both $q=3$ and $q=4$. Following this, after $T_{\text{c}}$, a minimum is observed in the variation of $d\langle G \rangle / dT$ with temperature, as shown in Fig.~\ref{fig:PEM34} (c) and (d). The temperature at which this minimum occurs corresponds to the system's third-order transition temperature, and these results align with those obtained through microcanonical inflection-point analysis \cite{wang2024exploring}.

\begin{figure*}[htbp]
    \centering
    \begin{minipage}{0.32\textwidth}
        \centering
        \epsfig{figure=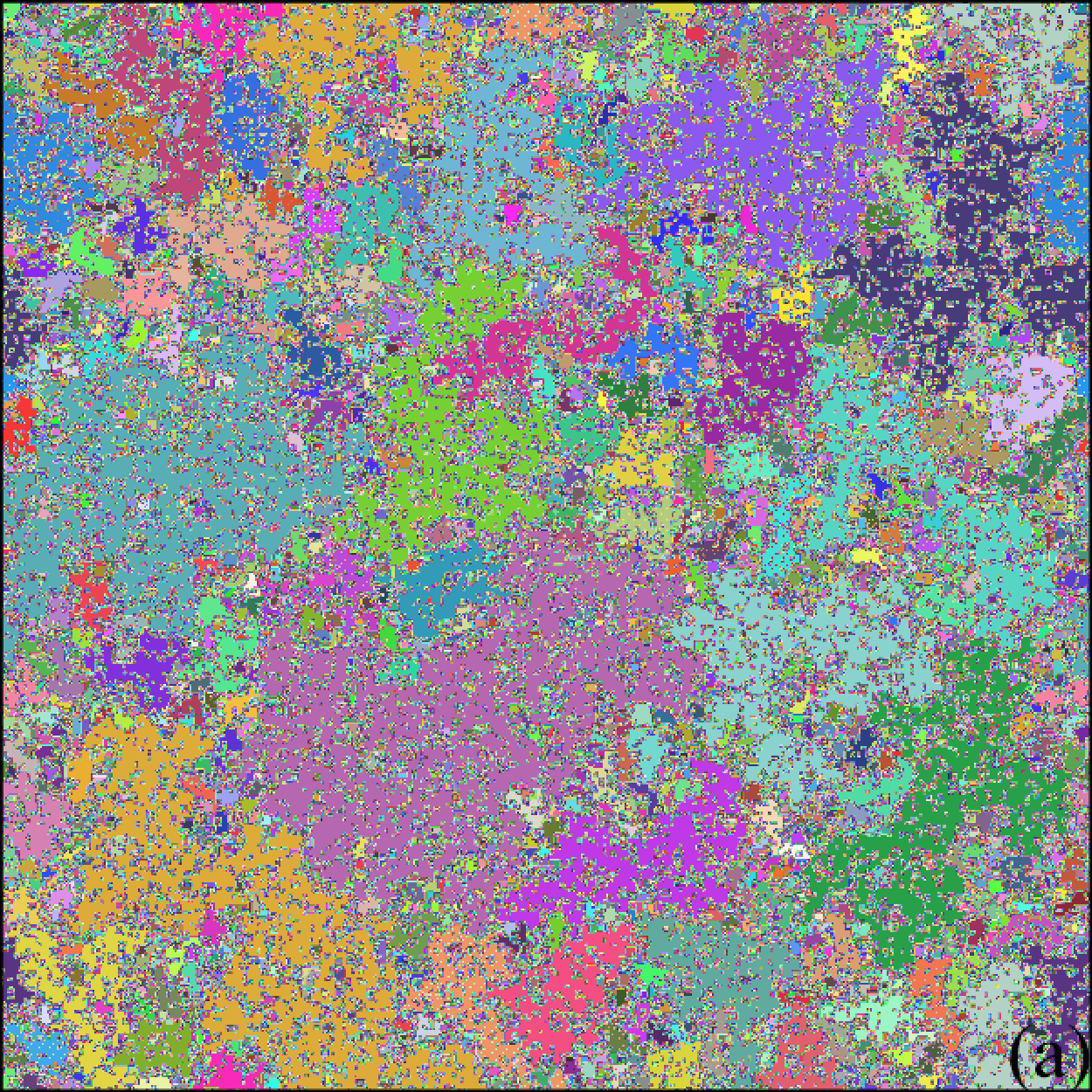,width=\linewidth}
        \label{fig:spin03_iso}
    \end{minipage} \hfill
    \begin{minipage}{0.32\textwidth}
        \centering
        \epsfig{figure=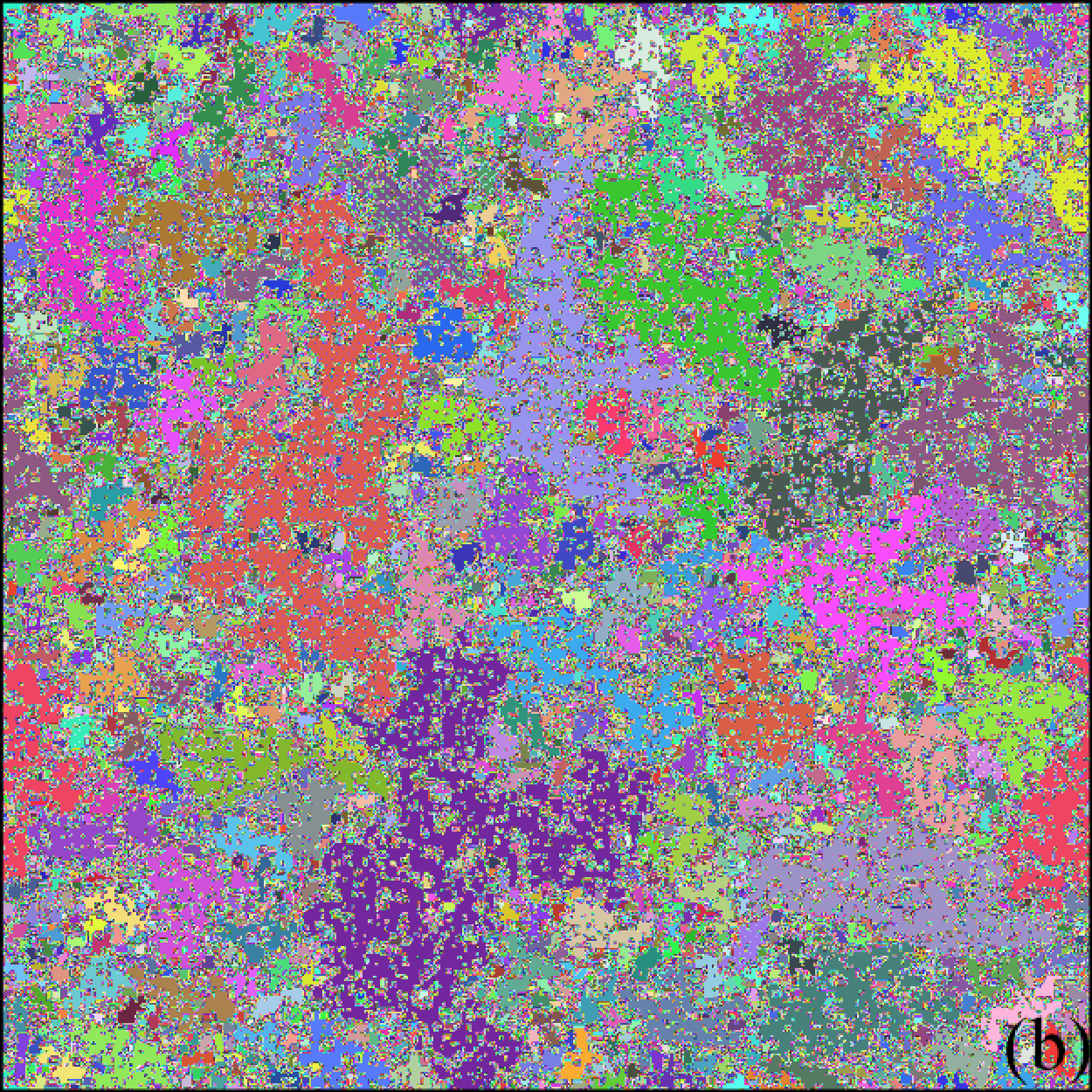,width=\linewidth}
        \label{fig:spin03_single}
    \end{minipage} \hfill
    \begin{minipage}{0.32\textwidth}
        \centering
        \epsfig{figure=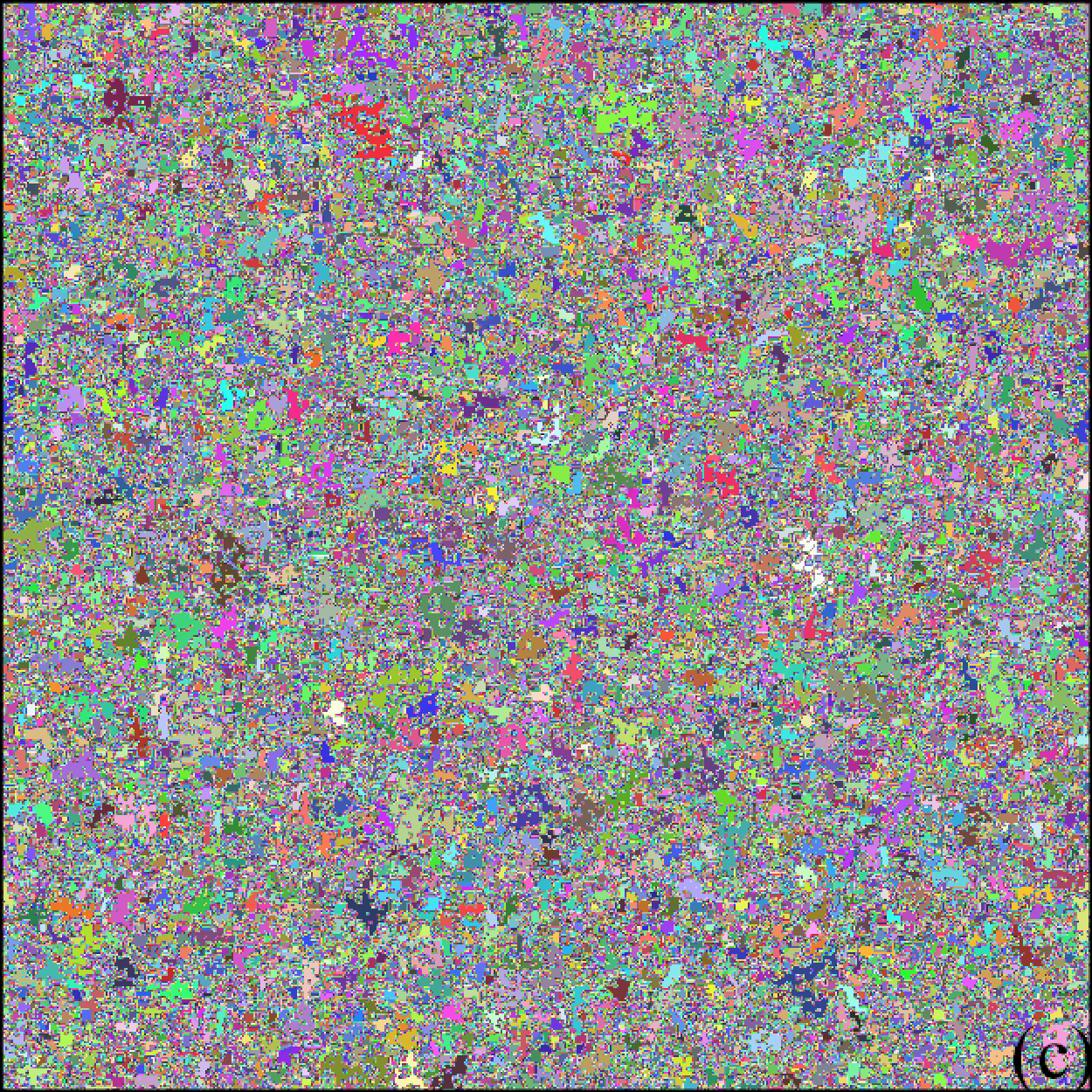,width=\linewidth}
        \label{fig:spin03_cluster}
    \end{minipage}
    \caption{
    Clusters identified in typical spin configurations on the 500 × 500 lattice for $q=3$ at three different temperatures: (a)$T \approx 0.996 $, is the phase transition temperature. (b)\( T \approx 1.000 \), just above the phase transition temperature, is approaching third dependent transition temperature. (c)$T \approx 1.060 $, is the third dependent transition temperature.
    }
    \label{fig:spin03_all}
\end{figure*}

By comparing the average perimeter results for the three-state and four-state Potts models in Fig.~\ref{fig:PEM34} (a) and (b), we observe a noteworthy phenomenon: the average perimeter $\langle G \rangle$ exhibits a local minimum prior to the phase transition and a local maximum subsequent to the transition. The maximum is referred to as the "peak," whereas the minimum is designated as the "valley." The difference between the peak and the valley is termed the "height difference." It is evident that as $q$ increases in the $q$-state Potts model (for $q \geq 3$), this height difference diminishes significantly. This observation reflects the behavior of clusters within the system under varying $q$-state conditions. As $q$ increases, the boundary conditions of the clusters become smoother, resulting in a simplification of the fractal structure. In Fig.~\ref{fig:PEM34} (c) and (d), it is observed that  $d\langle G \rangle / dT$ exhibits a local minimum. The temperature corresponding to this minimum is identified as the temperature $T_{de}$ associated with the third-order dependent transition. This minimum indicates that, within a temperature range $T$ prior to $T_{de}$, the average perimeter decreases as a concave function with increasing temperature; conversely, after $T_{de}$, it decreases as a convex function. Therefore, $T_{de}$ marks the inflection point at which the concavity of the average perimeter as a function of temperature changes, signifying the location of the third-order dependent transition. To gain a more intuitive understanding of the cluster information within the system, we simulated the clusters in the $q=3$ Potts model at three distinct temperatures: $T_{\text{c}}(3) \approx 0.996$, $T(3) \approx 1.000$, and $T_{de}(3) \approx 1.060$, as shown in Fig.~\ref{fig:spin03_all}.
\begin{figure*}[!htb]
    \centering  
    \includegraphics[width=1\textwidth]{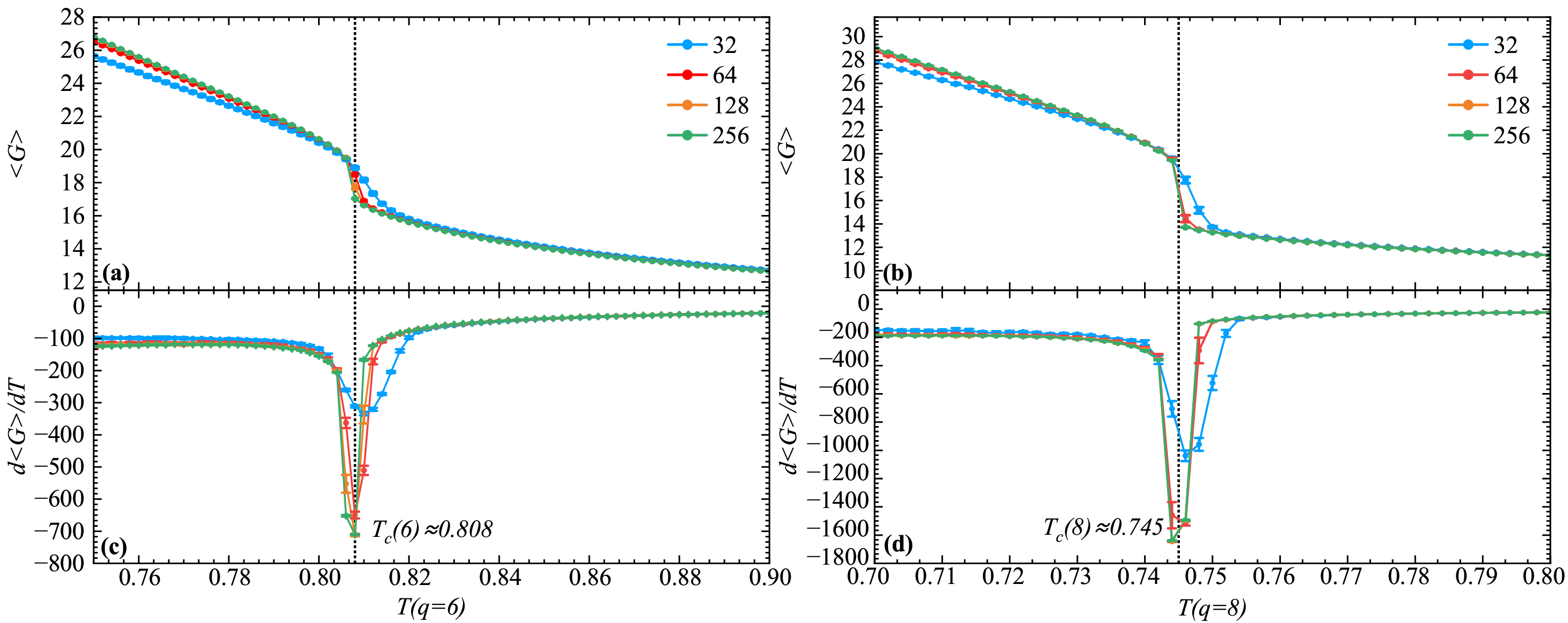} 
    \caption{Panels (a) and (b) show the variation of the average perimeter $\langle G \rangle$ with temperature for $q=6, 8$, while panels (c) and (d) depict the first derivative of the average perimeter with respect to temperature, $d\langle G \rangle / dT$. The black dashed line represents the exact solution for the phase transition temperature of the corresponding $q$-state Potts model. For any value of $q$, it is observed that as $L$ increases, the location of the minimum in $d\langle G \rangle / dT$ approaches the phase transition temperature. This minimum is not a signal of a third-order dependent transition, but rather indicates that as $q$ increases, the average perimeter $\langle G \rangle$ decreases with temperature, lacking any additional phase transition signals.
}
    \label{fig:PEM68}
\end{figure*}
In Fig.~\ref{fig:PEM68} (a) and (b), it is observed that the average perimeter $\langle G \rangle$ decreases as the temperature increases. Although the first derivative $d\langle G \rangle/dT$ exhibits a minimum, as shown in Fig.~\ref{fig:PEM68} (c) and (d), it is evident that this minimum shifts closer to the system's phase transition temperature $T_{\text{c}}$ as the system size increases, without indicating a signal of a third-order dependent transition. Previously, the order of the phase transition was determined by analyzing the relationship between $\langle P_{\infty} \rangle$ and $\langle M \rangle$, as shown in Fig.~\ref{fig:LCS}. For $q=6$ and $q=8$, the system undergoes a first-order phase transition. Based on this, it is speculated that no third-order dependent  transition signals exist in systems undergoing first-order transitions. In contrast, for $q=3$ and $q=4$, where the system exhibits a continuous phase transition, third-order dependent transition signals are present.

\section{Conclusion}
This study establishes the existence of third-order transitions in the Potts model from a geometric perspective and identifies the positions of both the third-order independent and dependent transitions using two order parameters: isolated spins number and the average perimeter. Comparisons between the microcanonical inflection points and transition positions determined by specific heat under canonical conditions show strong agreement. Furthermore, by analyzing the relationship between the Largest Cluster Size (LCS) and the magnetization $\langle M \rangle$, the order of phase transitions was identified for different values of $q$. The results show that for \( q = 3 \) and \( q = 4 \), the system undergoes continuous (second-order) transitions, whereas for \( q = 6 \) and \( q = 8 \), it exhibits discontinuous (first-order) transitions.

The number of isolated spins in the ordered phase usually reaches a peak, regardless of the value of \( q \). This indicates that, independent of the order of the system's main phase transition, a third-order independent transition usually exists, serving as a precursor to the impending disruption of the highly ordered state of the system. In the disordered phase of the system, we observed an inflection point in the rate of change of the average cluster perimeter, which corresponds to the existence of a third-order dependent transition. However, when the system undergoes a first-order phase transition, the dramatic changes in the clusters result in the absence of any higher-order phase transition signal in the system's average cluster perimeter in the disordered phase.

Based on the current research findings, we can confirm that the third-order independent transition usually exists, regardless of the order of the system's main phase transition. However, for the third-order dependent transition, when the phase transition from the ordered phase to the disordered phase is second-order, a third-order dependent transition is present in the system. But when the phase transition from the ordered phase to the disordered phase is first-order, the signal of the third-order dependent transition vanishes. Therefore, when the system undergoes a first-order main phase transition, whether a higher-order dependent transition exists in the disordered phase remains to be further studied and verified.

\section{Acknowledgments}
I would like to express my sincere gratitude to Professor Ying Tang for the invaluable discussions that significantly enriched this work and provided profound insights throughout the research process. This work is supported by the National Natural Science Foundation of China (Grant No.12304257).

\nocite{*}

\bibliography{referernces}

\end{document}